\documentclass[Physsubmission, Phys]{SciPost}

\hypersetup{
    colorlinks,
    linkcolor={red!50!black},
    citecolor={blue!50!black},
    urlcolor={blue!80!black}
}

\usepackage[bitstream-charter]{mathdesign}
\usepackage{caption}
\usepackage{subcaption}
\DeclareSymbolFont{usualmathcal}{OMS}{cmsy}{m}{n}
\DeclareSymbolFontAlphabet{\mathcal}{usualmathcal}

\newcommand\as{\alpha_{\mathrm{S}}}

\begin{document}

\begin{center}{\Large \textbf{Mixed QCD--EW corrections to $\boldsymbol{pp\!\to\!\ell\nu_\ell\!+\!X}$ at the LHC
\\
}}\end{center}

\begin{center}
Luca Buonocore\textsuperscript{1$\star$}
\end{center}

\begin{center}
{\bf 1} Physik Institut, Universit\"at Z\"urich, CH-8057 Z\"urich, Switzerland
\\
* lbuono@physik.uzh.ch
\end{center}

\begin{center}
\today
\end{center}


\definecolor{palegray}{gray}{0.95}
\begin{center}
\colorbox{palegray}{
  \begin{tabular}{rr}
  \begin{minipage}{0.1\textwidth}
    \includegraphics[width=35mm]{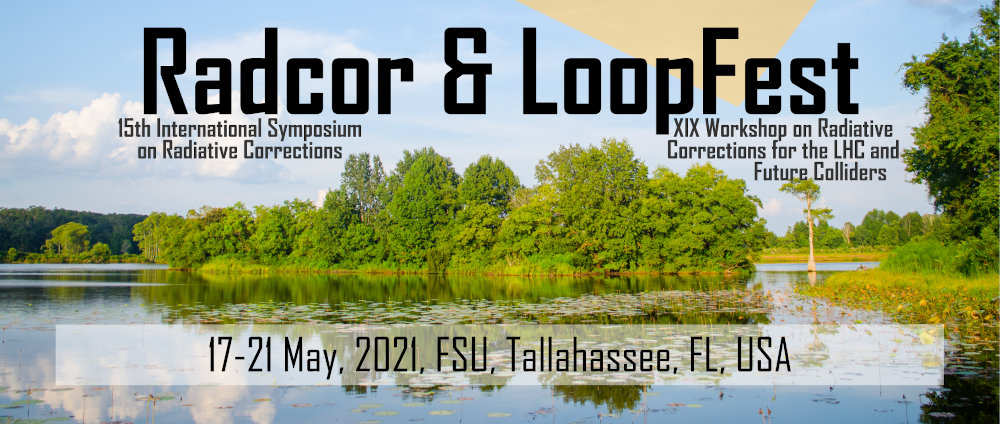}
  \end{minipage}
  &
  \begin{minipage}{0.85\textwidth}
    \begin{center}
    {\it 15th International Symposium on Radiative Corrections: \\Applications of Quantum Field Theory to Phenomenology,}\\
    {\it FSU, Tallahasse, FL, USA, 17-21 May 2021} \\
    \doi{10.21468/SciPostPhysProc.?}\\
    \end{center}
  \end{minipage}
\end{tabular}
}
\end{center}

\section*{Abstract}
{\bf
  We discuss about a recent computation of the mixed QCD--EW corrections to the hadroproduction
  of a massive charged lepton plus the corresponding neutrino through the Drell--Yan mechanism.
  The calculation, based on an extension of the $q_T$ subtraction formalism for heavy-quark production in
  next-to-next-to leading order QCD, includes for the first time all the real and virtual contributions
  due to initial- and final-state radiation,
  except for the finite part of the two-loop virtual correction, which is approximated
  in the pole approximation. We report results for the fiducial cross section and the transverse
  momentum spectrum of the charged lepton.
}

\vspace{10pt}
\noindent\rule{\textwidth}{1pt}
\tableofcontents\thispagestyle{fancy}
\noindent\rule{\textwidth}{1pt}
\vspace{10pt}

\section{Introduction}
\label{sec:intro}

The production of a dilepton pair via the Drell-Yan (DY) mechanism has
a special place in the precision phenomenology program at the LHC. Its
relatively large production rates and clean signatures, given the
presence of at least one charged lepton in the final state, make it a
standard candle process for benchmarking and experimental
calibrations.  DY data are of great importance for PDF fits and for
the (ultra)-precise measurement of fundamental electro-weak (EW)
parameters as the $W$ mass. Furthermore, this process represents an
important background for many New Physics searches.

Higher-order radiative corrections, both in strong and electro-weak
interactions, are mandatory to match the accuracy of experimental
results, expected to reach the (sub)percent level.  Theoretically, the
DY process is one of the most studied and well known processes: the
state-of-the-art for fully differential predictions is represented by
NNLO in QCD
~\cite{Anastasiou:2003yy,Anastasiou:2003ds,Melnikov:2006kv,Catani:2009sm,Catani:2010en},
and NLO in
EW~\cite{Dittmaier:2001ay,Baur:2004ig,Zykunov:2006yb,Arbuzov:2005dd,CarloniCalame:2006zq,Baur:2001ze,Zykunov:2005tc,CarloniCalame:2007cd,Arbuzov:2007db,Dittmaier:2009cr}.
Recently, N$^3$LO QCD radiative corrections of the inclusive
production of a virtual photon~{\cite{Duhr:2020seh,Chen:2021vtu}} and
of a $W$ boson~\cite{Duhr:2020sdp} have been computed, and first
estimates of fiducial cross sections for the neutral-current DY
process at the same order have appeared~\cite{Camarda:2021ict}.

In view of the level of precision attainable at the LHC, it becomes
relevant to include ${\cal O}(\as\alpha)$ mixed QCD--EW
corrections. We discuss the case of the charged-current DY process
\begin{equation}
  \label{eq:proc}
pp\to \ell^+\nu_\ell+X.
\end{equation}
The computation of mixed corrections is a complicated task, the
complexity being that of a NNLO calculation for a $2\to2$ process with
many scales. One of the bottlenecks is the corresponding two-loop
virtual amplitude. The evaluation of the $2\to 2$ two-loop Feynman
diagrams with internal masses is at the frontier of current
computational techniques and the corresponding amplitude has not yet
been fully worked out.

We present results for the mixed corrections to the charged current
process, including, for the first time, all real and virtual
contributions. Everything is exact but the two-loop virtual amplitude,
which is approximated by its expansion around the resonant pole,
applying the Pole Approximation of Ref.~\cite{Dittmaier:2014qza}.

\section{Structure of the computation}
The differential cross section for the process in Eq.~(\ref{eq:proc})
can be written as
\begin{equation}
  \label{eq:exp}
  d{\sigma}=\sum_{m,n=0}^\infty d{\sigma}^{(m,n)}\, ,
\end{equation}
where $d{\sigma}^{(0,0)}\equiv d{\sigma}_{\rm LO}$ is the Born level
contribution and $d{\sigma}^{(m,n)}$ the ${\cal O}(\as^m\alpha^n)$
correction.  The mixed QCD--EW corrections correspond to the term
$m=n=1$ in this expansion.

We achieve the cancellation of the infrared divergences by exploiting
the $q_T$ subtraction formalism~\cite{Catani:2007vq}. The extension of
the method to the NLO EW and the mixed QCD--EW corrections have been worked
out in Refs.~\cite{Buonocore:2019puv} and~\cite{Buonocore:2021rxx},
starting from the $q_T$ subtraction formalism for heavy
quarks~\cite{Catani:2019iny}.

In the following we denote the transverse momentum of the system
formed by the charged lepton and its corresponding neutrino with
$q_T$. Focusing on the case of the mixed corrections, we have
schematically that $d{\sigma}^{(1,1)}$ can be evaluated as
\begin{equation}
  \label{eq:master}
  d{\sigma}^{(1,1)}={\cal H}^{(1,1)}\otimes d{\sigma}_{\rm LO}+\left[d\sigma_{\rm R}^{(1,1)}-d\sigma_{\rm CT}^{(1,1)}\right]\,
\end{equation}
where
\begin{itemize}
  \item $d\sigma_{\rm R}^{(1,1)}$ is the {\it real} contribution
    associated to configurations in which the charged lepton and the
    corresponding neutrino are accompanied by additional QCD and/or
    QED radiation that produces a recoil with finite transverse
    momentum $q_T$;
  \item the customary $q_T$ counterterm $d\sigma_{\rm CT}^{(1,1)}$  
    cancels the singular behaviour in the limit $q_T\to 0$,
    rendering the cross section in Eq.~(\ref{eq:master})
    finite;
  \item ${\cal H}^{(1,1)}$ is a perturbatively computable function
    which encodes the contribution stemming from the two-loop virtual
    amplitude.
\end{itemize}

The symbol $\otimes$ in the first term of Eq.~(\ref{eq:master})
denotes a convolution with respect to the longitudinal-momentum
fractions $z_1$ and $z_2$ of the colliding partons.

In particular, the coefficient ${\cal H}^{(1,1)}$ can be decomposed as
\begin{equation}
  {\cal H}^{(1,1)}=H^{(1,1)}\delta(1-z_1)\delta(1-z_2)+\delta{\cal H}^{(1,1)}\,,
\end{equation}
where the hard contribution $H^{(1,1)}$ contains the 2-loop virtual
corrections.  More precisely, the finite contribution is defined as
\begin{equation}
  \label{eq:H11def}
  H^{(1,1)}=\frac{2{\rm Re}\left({\cal M}_{\rm fin}^{(1,1)}{\cal
      M}^{(0,0)*}\right)}{|{\cal M}^{(0,0)}|^2} \, .
\end{equation}
in terms of the Born amplitude ${\cal M}^{(0,0)}$ and of the finite
part, ${\cal M}_{\rm fin}^{(1,1)}$, of the renormalised virtual
amplitudes ${\cal M}^{(1,1)}$ entering the mixed QCD--EW calculations,
after subtraction of the infrared poles in $d=4-2\epsilon$ dimensions
with a customary subtraction operator~\cite{Buonocore:2021rxx}.

\begin{figure}[t]
  \begin{center}
    \begin{subfigure}[b]{0.3\textwidth}    
      \includegraphics[width=\textwidth]{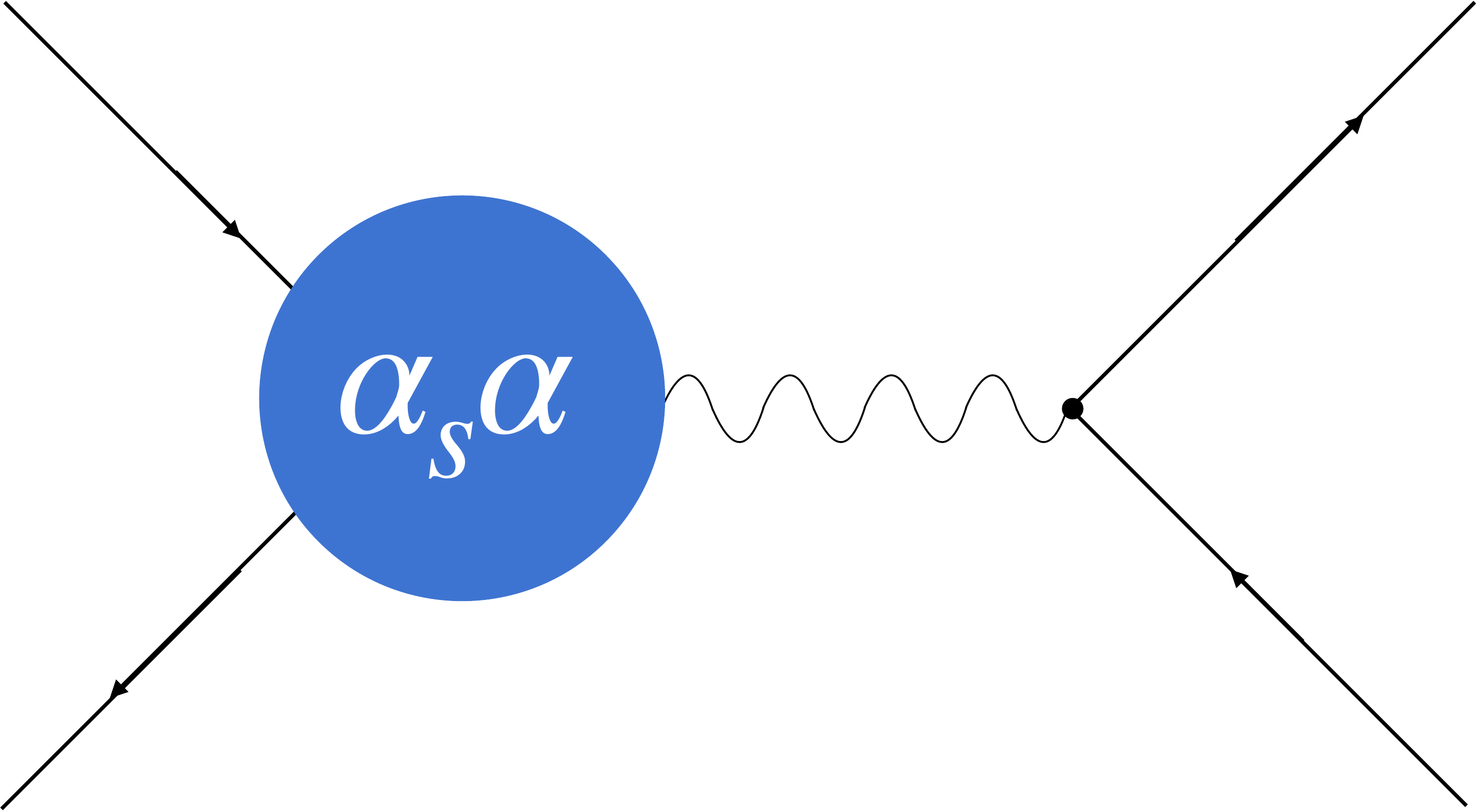}
      \caption{Factorizable, initial-initial.}
      \label{fig:PA-ii}
    \end{subfigure}
    \hfill
    \begin{subfigure}[b]{0.34\textwidth}    
      \includegraphics[width=\textwidth]{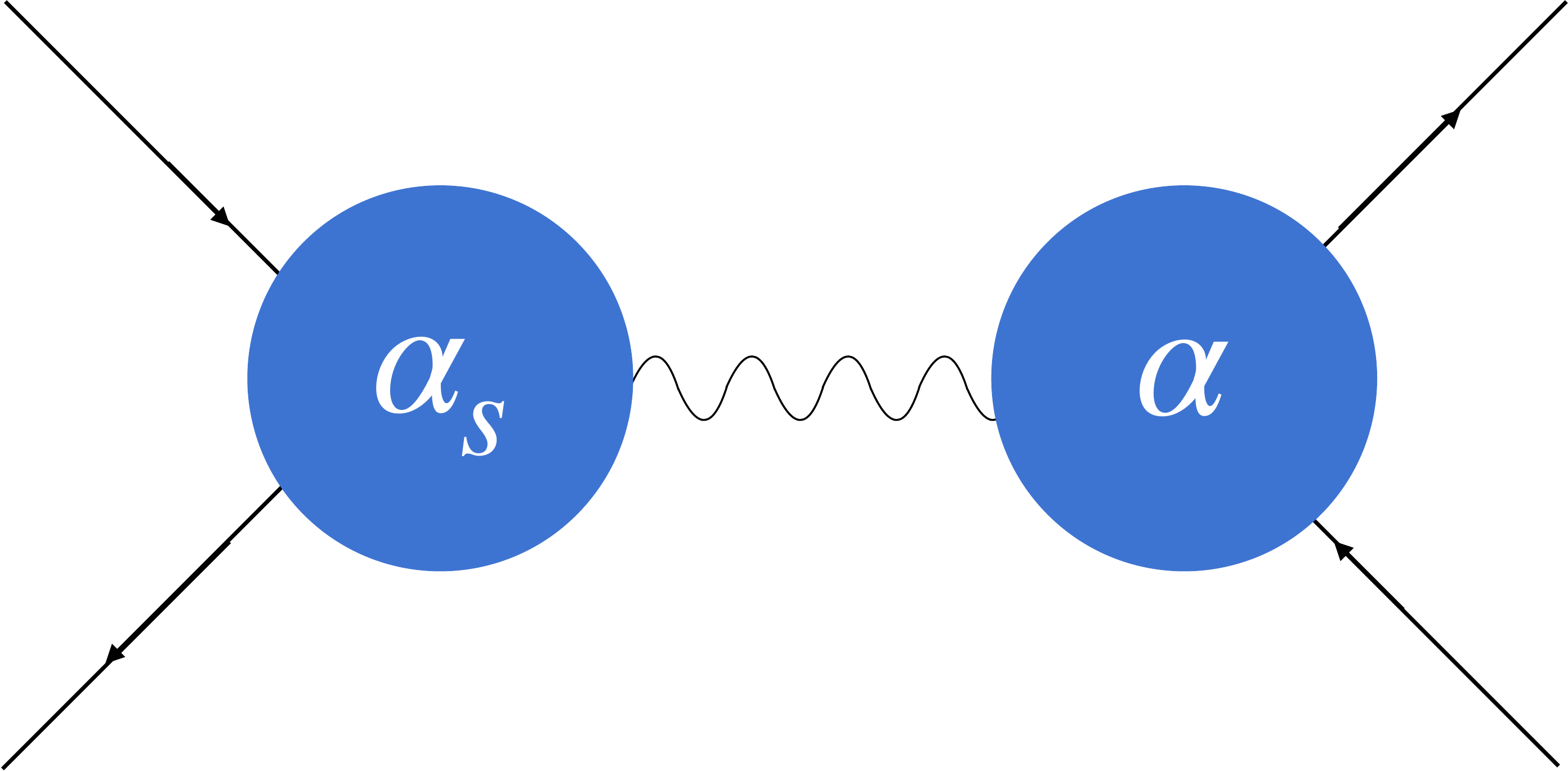}
      \caption{Factorizable, initial-final.}
      \label{fig:PA-if}
    \end{subfigure}
    \hfill
    \begin{subfigure}[b]{0.3\textwidth}    
      \includegraphics[width=\textwidth]{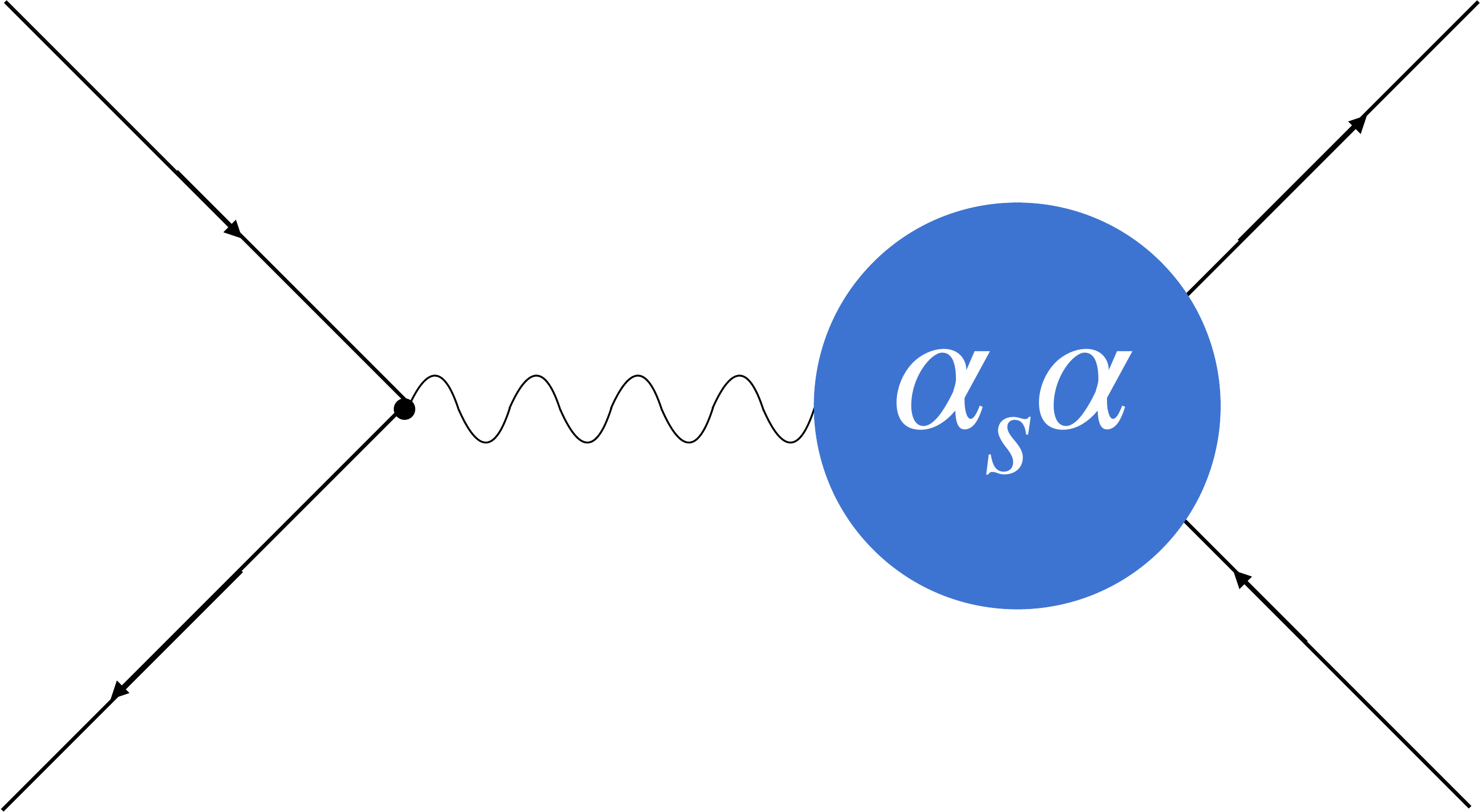}
      \caption{Factorizable, final-final.}
      \label{fig:PA-if}
    \end{subfigure}    
    \\
    \vspace{0.5cm}
    \begin{subfigure}[b]{0.34\textwidth}    
      \includegraphics[width=\textwidth]{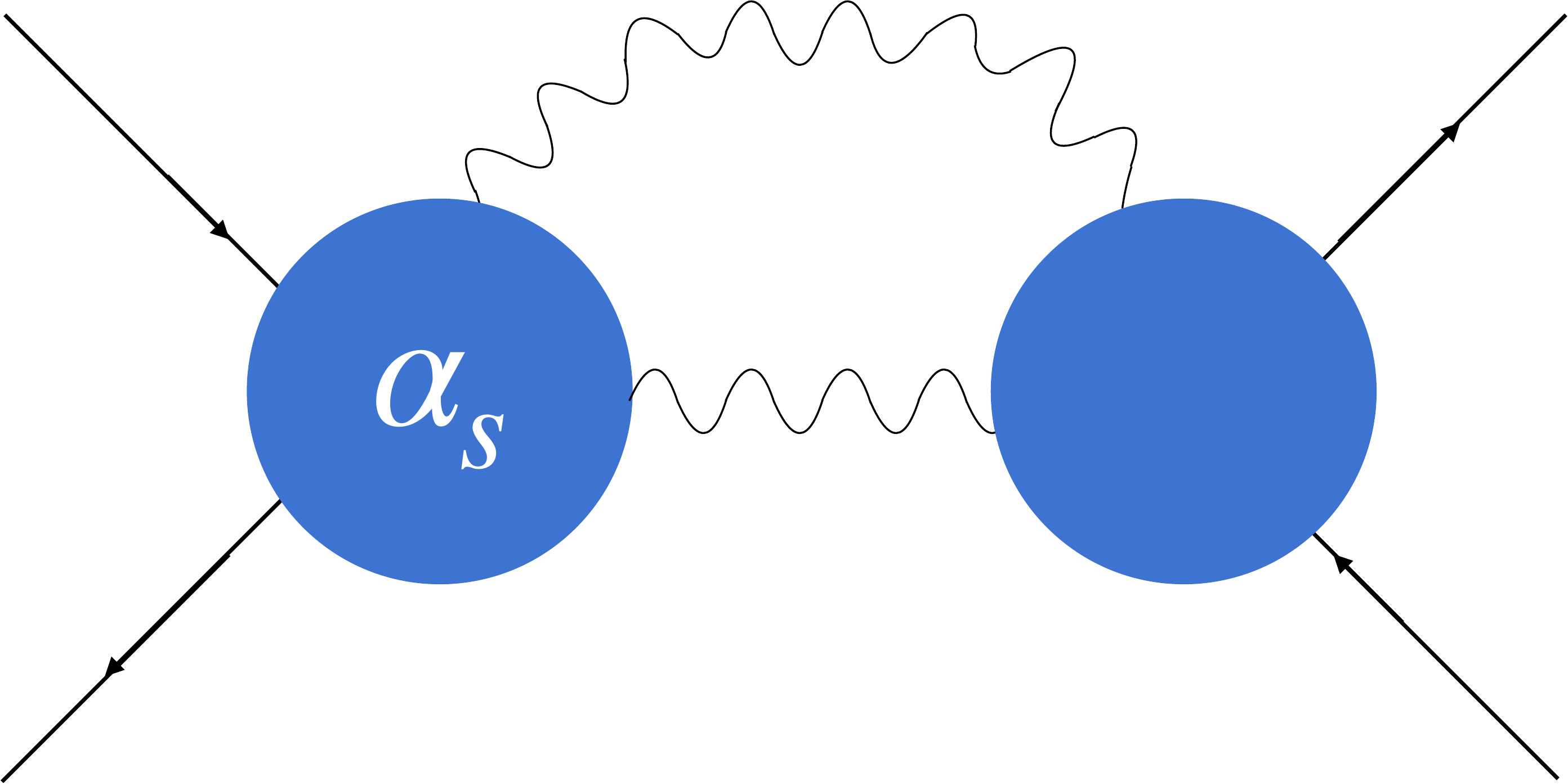}
      \caption{Non factorizable.}
      \label{fig:PA-nf}
    \end{subfigure}   
\end{center}
\vspace{-2ex}
\caption{\label{fig:PA}
  Factorizable and non factorizable contributions to the pole approximation. }
\end{figure}

For the computation ${\cal M}_{\rm fin}^{(1,1)}$, we approximate the
renormalised virtual amplitudes ${\cal M}^{(1,1)}$ in pole
approximation including all the factorizable and non-factorizable
contributions, see Fig.\ref{fig:PA}.  In particular, we include the
initial-intial factorizable contribution, Fig.\ref{fig:PA-ii} via the $W$ boson two-loop
form factor of Ref.~\cite{Behring:2020cqi}.

We remark that the renormalised amplitude in PA, ${\cal M}_{\rm
  PA}^{(1,1)}$, presents the same structure of poles as the exact one;
thereby, the finite part is consistently extracted applying the same
subtraction operator. 

The quality of the pole approximation for the charged current
Drell-Yan process has been discussed in Ref.~\cite{Buonocore:2021rxx},
based on the comparison with the exact results at the NLO
level. Furthermore, the same work showed that the pole approximation
can be improved beyond the resonant region by applying suitable
reweighting factors. Recently, the exact mixed QCD--EW corrections to
the neutral current process have been
computed~\cite{Bonciani:2021zzf}, providing a direct comparison
between the exact two-loop amplitude and its pole approximation.  In
the following section, we will show result for the charged current
process adopting the reweighting prescription of
Ref.~\cite{Buonocore:2021rxx}.

Before concluding this section, we mention that all the remaining real
and virtual ${\cal O}(\as\alpha)$ contributions are evaluated without
any approximation using {\sc Openloops}~\cite{Cascioli:2011va,
  Buccioni:2017yxi, Buccioni:2019sur} and {\sc
  Recola}~\cite{Actis:2016mpe,Denner:2017wsf}. The required phase
space generation and integration is carried out within the {\sc
  Matrix} framework~\cite{Grazzini:2017mhc}.

\section{Results}
\label{sec:results}
We consider the process $pp\to \mu^+\nu_\mu+X$ at centre-of-mass energy \mbox{$\sqrt{s}=14$\,TeV}. We adopt a setup similar to Ref.~\cite{Dittmaier:2015rxo}, and, in particular, we work in the $G_\mu$ scheme with
\begin{align}
  G_F &=1.1663787\times 10^{-5}~{\rm GeV}^{-2}  &\alpha(0)&=1/137.03599911\\
  m_{W, {\rm OS}}&=80.385~{\rm GeV}             &m_{Z, {\rm OS}}&=91.1876~{\rm GeV}\\
  \Gamma_{W, {\rm OS}}&=2.085~{\rm GeV}        &\Gamma_{Z, {\rm OS}}&=2.4952~{\rm GeV}\\
  m_H&=173.07~{\rm GeV}                    &m_H&=125.9~{\rm GeV}\\
  \text{pdf: }& \texttt{NNPDF31$\_$nnlo$\_$as$\_$0118$\_$luxqed}   &\mu_F&=\mu_R = m_W.
\end{align}
We use the complex-mass scheme~\cite{Denner:2005fg} throughout and a diagonal CKM matrix. \\ The \textit{on-shell} values are translated to the corresponding \textit{pole} values $m_{V}=m_{V,{\rm OS}}/\sqrt{1+\Gamma^2_{V,{\rm OS}}/m^2_{V{,\rm OS}}}$ and $\Gamma_{V}=\Gamma_{V,{\rm OS}}/\sqrt{1+\Gamma^2_{V,{\rm OS}}/m_{V,{\rm OS}}^2}$, $V=W,Z$, from which  $\alpha=\sqrt{2}\,G_F m_{W}^2(1-m_{W}^2/m_{Z}^2)/\pi$ is derived. The muon mass is fixed to $m_\mu=105.658369$\,MeV.
We use the following selection cuts,
\begin{equation}
  \label{eq:cuts}
  p_{T,\mu}>25\,{\rm GeV}\,,\qquad |y_\mu|<2.5\,,\qquad p_{T,\nu}>25\,{\rm GeV}\,,
\end{equation}
and work at the level of {\it bare muons}, i.e., no lepton recombination with close-by photons is carried out.

\subsection{Fiducial cross section}
In Tab.~\ref{tab:fid} we present our predictions for the fiducial cross section corresponding to the selection cuts in Eq.~\eqref{eq:cuts}. We show the breakdown of the different contributions $\sigma^{(i,j)}$ into the various partonic channels. 
The contribution from the channels $u{\bar d},\,c{\bar s}$ is denoted by $q{\bar q}$.
\begin{table}
\renewcommand{\arraystretch}{1.4}
  \centering
  \begin{tabular}{|c|c|c|c|c|c|}
    \hline
    $\sigma$ [pb] & $\sigma_{\rm LO}$  & $\sigma^{(1,0)}$  & $\sigma^{(0,1)}$  & $\sigma^{(2,0)}$ & $\sigma^{(1,1)}$ \\
    \hline
    \hline
    $q{\bar q}$ & $5029.2$ & $\phantom{+0}970.5(3)\phantom{00}$ & $-143.61(15)$ & $\phantom{+}251(4)\phantom{.0}$ & $\hspace*{-0.6ex}\phantom{0}-7.0(1.2)\phantom{00}\hspace*{-0.6ex}$\\
    \hline
    $qg$ & --- & $-1079.86(12)$ & --- & $-377(3)\phantom{.0}$ & $\hspace*{-0.6ex}\phantom{+}39.0(4)\phantom{.00}\hspace*{-0.6ex}$ \\
    \hline
    $q(g)\gamma$ & --- & --- & $\phantom{+00}2.823(1)$ & --- & $\hspace*{-0.6ex}\phantom{+0}0.055(5)\phantom{.0}\hspace*{-0.6ex}$\\
    \hline
    $q({\bar q})q^\prime$ & --- & --- & --- & $\phantom{+0}44.2(7)$ & $\hspace*{-0.6ex}\phantom{+0}1.2382(3)\phantom{.}\hspace*{-0.6ex}$\\
    \hline
    $g g$ & --- & --- & --- & $\phantom{+}100.8(8)$ & --- \\
    \hline
    \hline
    tot & $5029.2$ & $\phantom{0}$$-109.4(4)\phantom{00}$ & $-140.8(2)\phantom{00}$ & $\phantom{00}19(5)\phantom{.0}$ &  $\hspace*{-0.6ex}\phantom{+} 33.3(1.3)\phantom{00}\hspace*{-0.6ex}$ \\
    \hline
    \hline
    $\sigma/\sigma_{\rm LO}$ & $1$ & $-2.2\%$ &  $-2.8\%$ &  $+0.4\%$  &   $+0.6\%$  \\
    \hline  \end{tabular}
  \caption{\label{tab:fid}
    The different perturbative contributions to the fiducial cross section and their breakdown into the various partonic channels is also shown. The numerical uncertainties are stated in brackets.}
\renewcommand{\arraystretch}{1.0}
  \end{table}
The contributions from the channels
$qg,\,{\bar q}g$ and $q\gamma,\,{\bar q}\gamma$, which enter at NLO QCD and EW, are labelled by $qg$ and $q\gamma$, respectively.
The contribution from all the remaining quark--quark channels
$qq',\, {\bar q}{\bar q}',\, q{\bar q}'$ (excluding $u{\bar d},\,c{\bar s}$) to the NNLO QCD and mixed corrections
is labelled by $q({\bar q})q^\prime$.
Finally, the contributions from the gluon--gluon and gluon--photon channels, which are relevant only at ${\cal O}(\as^2)$ and ${\cal O}(\as\alpha)$, are denoted by $gg$ and $g\gamma$, respectively.

We see that
\begin{itemize}
\item (N)NLO QCD corrections are subject to large cancellations between the different partonic channels; for this reason, 
  NNLO QCD and mixed QCD-EW corrections, as well as NLO QCD and NLO EW ones, have a similar quantitative impact;
\item mixed QCD-EW corrections are positive and dominated by the $qg$ channel, which is exact in our calculation. 
\end{itemize}
Because of the large cancellations occurring for central scales, we observe that the pattern of the higher-order QCD corrections to the perturbative series has a strong dependence on the choice of the renormalisation and factorisation scales. 
For example, the scale choice $\mu_R=\mu_F={m_W}/{2}$ leads to a more common perturbative pattern: \\
$\sigma^{(1,0)}/\sigma_{\rm LO} = +10\%$, $\sigma^{(0,1)}/\sigma_{\rm LO} = -2.9\%$, $\sigma^{(2,0)}/\sigma_{\rm LO} =+4.2\%$, $\sigma^{(1,1)}/\sigma_{\rm LO} = +0.76\%$.
\subsection{Transverse momentum spectrum $p_{T,\mu^+}$}
\begin{figure}[t]
\begin{center}
\includegraphics[width=0.46\textwidth]{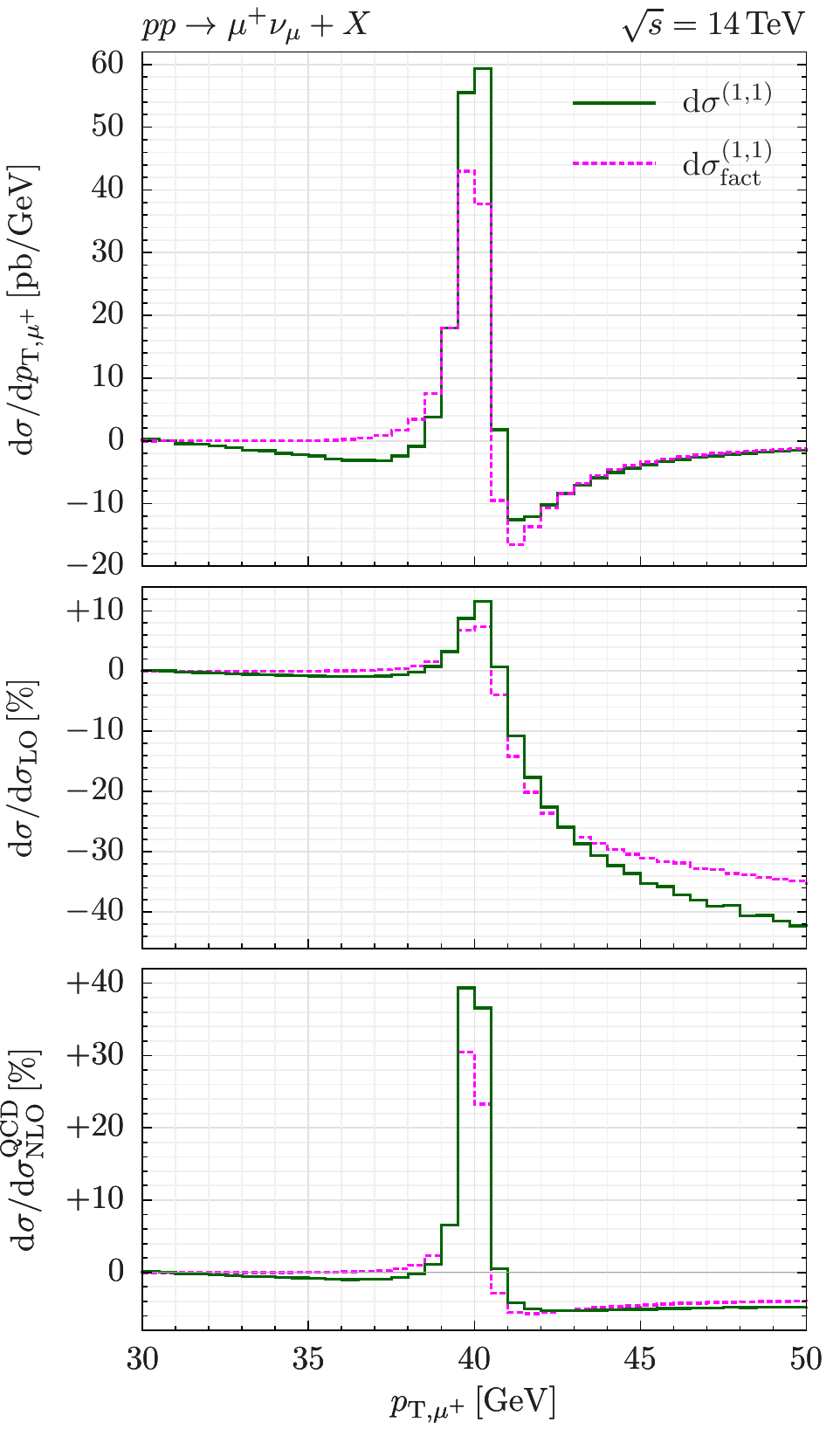}\hfill
\includegraphics[width=0.46\textwidth]{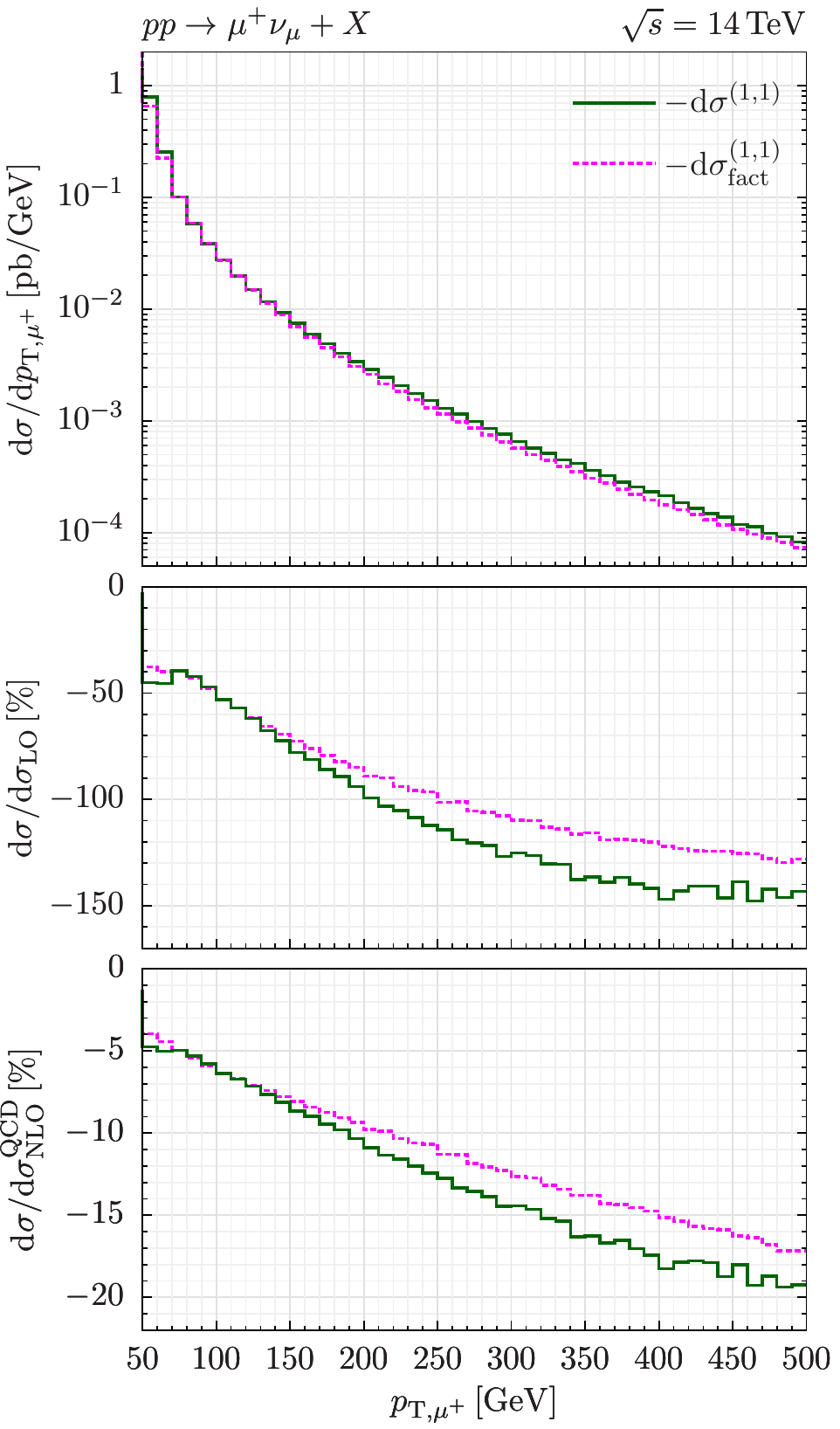}\\
\end{center}
\vspace{-2ex}
\caption{\label{fig:ptOS-OFFS}
Complete ${\cal O}(\as\alpha)$ correction to the differential cross section $d\sigma^{(1,1)}$ in the muon $p_T$, and its factorized approximation $d\sigma^{(1,1)}_{\rm fact}$, defined in Eq.~(\ref{eq:fact}).
The top panels show the absolute predictions, while the central (bottom) panels display the ${\cal O}(\as\alpha)$ correction normalized to the LO (NLO QCD) result.
}
\end{figure}

In Fig.~\ref{fig:ptOS-OFFS}, we show our results for the complete ${\cal O}(\as\alpha)$ correction to the transverse momentum spectrum of the positively charged muon. 
We compare our results with those obtained by a multiplicative combination of the NLO QCD and NLO EW corrections. The latter approach is justified under the assumption of a completely factorisation of the two corrections. We define the multiplicative combination as follows:
for each bin, the QCD correction, $d\sigma^{(1,0)}/dp_T$, and the EW correction restricted to the $q{\bar q}$ channel,
$d\sigma_{q{\bar q}}^{(0,1)}/dp_T$, are computed, and the factorised ${\cal O}(\as\alpha)$ correction is calculated as
\begin{equation}
  \label{eq:fact}
\frac{d\sigma^{(1,1)}_{\rm fact}}{dp_T}=\left(\frac{d\sigma^{(1,0)}}{dp_T}\right)\times\left(\frac{d\sigma_{q{\bar q}}^{(0,1)}}{dp_T}\right)\times\left(\frac{d\sigma_{\rm LO}}{dp_T}\right)^{-1}\, ,
\end{equation}
We observe that the factorised approximation reproduces qualitatively well our
result for the ${\cal O}(\as\alpha)$. Beyond the Jacobian peak, it
tends to overshoot the complete result.
As $p_T$ increases, the negative impact of the mixed QCD--EW corrections increases and becomes rather sizeable, reaching at $p_T=500$\,GeV about $-140\%$ with respect to the LO prediction and $-20\%$ with respect to the NLO QCD result. This is not unexpected, since the high-$p_T$ region is dominated by $W+{\rm jet}$ topologies, for which the ${\cal O}(\as\alpha)$ effects can be seen as NLO EW corrections.

\section{Conclusion}

Higher-order mixed QCD-EW corrections are relevant for processes that can be measured at the (sub)percent level, as the production of a lepton pair through the Drell-Yan mechanism.  
In this contribution, we presented new results for $\cal{O}(\as\alpha)$ corrections to the hadroproduction of a massive charged lepton plus the corresponding neutrino at the LHC, showing their impact on fiducial cross sections and on the transverse momentum spectrum of the charged lepton.


\paragraph{Funding information}
This material is based upon work supported in part by the Swiss National Science Foundation (SNF) under contracts IZSAZ2$\_$173357 and 200020$\_$188464, by the ERC Starting Grant 714788 REINVENT and by INFN.



\bibliography{biblio.bib}

\begin{thebibliography}{10}
\providecommand{\url}[1]{\texttt{#1}}
\providecommand{\urlprefix}{URL }
\expandafter\ifx\csname urlstyle\endcsname\relax
  \providecommand{\doi}[1]{doi:\discretionary{}{}{}#1}\else
  \providecommand{\doi}{doi:\discretionary{}{}{}\begingroup
  \urlstyle{rm}\Url}\fi
\providecommand{\eprint}[2][]{\url{#2}}

\bibitem{Anastasiou:2003yy}
C.~Anastasiou, L.~J. Dixon, K.~Melnikov and F.~Petriello,
\newblock \emph{{Dilepton rapidity distribution in the Drell-Yan process at
  NNLO in QCD}},
\newblock Phys. Rev. Lett. \textbf{91}, 182002 (2003),
\newblock \doi{10.1103/PhysRevLett.91.182002},
\newblock \eprint{hep-ph/0306192}.

\bibitem{Anastasiou:2003ds}
C.~Anastasiou, L.~J. Dixon, K.~Melnikov and F.~Petriello,
\newblock \emph{{High precision QCD at hadron colliders: Electroweak gauge
  boson rapidity distributions at NNLO}},
\newblock Phys. Rev. D \textbf{69}, 094008 (2004),
\newblock \doi{10.1103/PhysRevD.69.094008},
\newblock \eprint{hep-ph/0312266}.

\bibitem{Melnikov:2006kv}
K.~Melnikov and F.~Petriello,
\newblock \emph{{Electroweak gauge boson production at hadron colliders through
  $\mathcal{O}(\alpha_s^2)$}},
\newblock Phys. Rev. D \textbf{74}, 114017 (2006),
\newblock \doi{10.1103/PhysRevD.74.114017},
\newblock \eprint{hep-ph/0609070}.

\bibitem{Catani:2009sm}
S.~Catani, L.~Cieri, G.~Ferrera, D.~de~Florian and M.~Grazzini,
\newblock \emph{{Vector boson production at hadron colliders: a fully exclusive
  QCD calculation at NNLO}},
\newblock Phys. Rev. Lett. \textbf{103}, 082001 (2009),
\newblock \doi{10.1103/PhysRevLett.103.082001},
\newblock \eprint{0903.2120}.

\bibitem{Catani:2010en}
S.~Catani, G.~Ferrera and M.~Grazzini,
\newblock \emph{{W Boson Production at Hadron Colliders: The Lepton Charge
  Asymmetry in NNLO QCD}},
\newblock JHEP \textbf{05}, 006 (2010),
\newblock \doi{10.1007/JHEP05(2010)006},
\newblock \eprint{1002.3115}.

\bibitem{Dittmaier:2001ay}
S.~Dittmaier and M.~Kr\"amer,
\newblock \emph{{Electroweak radiative corrections to W boson production at
  hadron colliders}},
\newblock Phys. Rev. D \textbf{65}, 073007 (2002),
\newblock \doi{10.1103/PhysRevD.65.073007},
\newblock \eprint{hep-ph/0109062}.

\bibitem{Baur:2004ig}
U.~Baur and D.~Wackeroth,
\newblock \emph{{Electroweak radiative corrections to $p \bar{p} \to W^\pm \to
  \ell^\pm \nu$ beyond the pole approximation}},
\newblock Phys. Rev. D \textbf{70}, 073015 (2004),
\newblock \doi{10.1103/PhysRevD.70.073015},
\newblock \eprint{hep-ph/0405191}.

\bibitem{Zykunov:2006yb}
V.~Zykunov,
\newblock \emph{{Radiative corrections to the Drell-Yan process at large
  dilepton invariant masses}},
\newblock Phys. Atom. Nucl. \textbf{69}, 1522 (2006),
\newblock \doi{10.1134/S1063778806090109}.

\bibitem{Arbuzov:2005dd}
A.~Arbuzov, D.~Bardin, S.~Bondarenko, P.~Christova, L.~Kalinovskaya, G.~Nanava
  and R.~Sadykov,
\newblock \emph{{One-loop corrections to the Drell-Yan process in SANC. I. The
  Charged current case}},
\newblock Eur. Phys. J. C \textbf{46}, 407 (2006),
\newblock \doi{10.1140/epjc/s2006-02505-y},
\newblock [Erratum: Eur.Phys.J.C 50, 505 (2007)],
\newblock \eprint{hep-ph/0506110}.

\bibitem{CarloniCalame:2006zq}
C.~Carloni~Calame, G.~Montagna, O.~Nicrosini and A.~Vicini,
\newblock \emph{{Precision electroweak calculation of the charged current
  Drell-Yan process}},
\newblock JHEP \textbf{12}, 016 (2006),
\newblock \doi{10.1088/1126-6708/2006/12/016},
\newblock \eprint{hep-ph/0609170}.

\bibitem{Baur:2001ze}
U.~Baur, O.~Brein, W.~Hollik, C.~Schappacher and D.~Wackeroth,
\newblock \emph{{Electroweak radiative corrections to neutral current Drell-Yan
  processes at hadron colliders}},
\newblock Phys. Rev. D \textbf{65}, 033007 (2002),
\newblock \doi{10.1103/PhysRevD.65.033007},
\newblock \eprint{hep-ph/0108274}.

\bibitem{Zykunov:2005tc}
V.~Zykunov,
\newblock \emph{{Weak radiative corrections to Drell-Yan process for large
  invariant mass of di-lepton pair}},
\newblock Phys. Rev. D \textbf{75}, 073019 (2007),
\newblock \doi{10.1103/PhysRevD.75.073019},
\newblock \eprint{hep-ph/0509315}.

\bibitem{CarloniCalame:2007cd}
C.~Carloni~Calame, G.~Montagna, O.~Nicrosini and A.~Vicini,
\newblock \emph{{Precision electroweak calculation of the production of a high
  transverse-momentum lepton pair at hadron colliders}},
\newblock JHEP \textbf{10}, 109 (2007),
\newblock \doi{10.1088/1126-6708/2007/10/109},
\newblock \eprint{0710.1722}.

\bibitem{Arbuzov:2007db}
A.~Arbuzov, D.~Bardin, S.~Bondarenko, P.~Christova, L.~Kalinovskaya, G.~Nanava
  and R.~Sadykov,
\newblock \emph{{One-loop corrections to the Drell--Yan process in SANC. (II).
  The Neutral current case}},
\newblock Eur. Phys. J. C \textbf{54}, 451 (2008),
\newblock \doi{10.1140/epjc/s10052-008-0531-8},
\newblock \eprint{0711.0625}.

\bibitem{Dittmaier:2009cr}
S.~Dittmaier and M.~Huber,
\newblock \emph{{Radiative corrections to the neutral-current Drell-Yan process
  in the Standard Model and its minimal supersymmetric extension}},
\newblock JHEP \textbf{01}, 060 (2010),
\newblock \doi{10.1007/JHEP01(2010)060},
\newblock \eprint{0911.2329}.

\bibitem{Duhr:2020seh}
C.~Duhr, F.~Dulat and B.~Mistlberger,
\newblock \emph{{Drell-Yan Cross Section to Third Order in the Strong Coupling
  Constant}},
\newblock Phys. Rev. Lett. \textbf{125}(17), 172001 (2020),
\newblock \doi{10.1103/PhysRevLett.125.172001},
\newblock \eprint{2001.07717}.

\bibitem{Chen:2021vtu}
X.~Chen, T.~Gehrmann, N.~Glover, A.~Huss, T.-Z. Yang and H.~X. Zhu,
\newblock \emph{{Di-lepton Rapidity Distribution in Drell-Yan Production to
  Third Order in QCD}}  (2021),
\newblock \eprint{2107.09085}.

\bibitem{Duhr:2020sdp}
C.~Duhr, F.~Dulat and B.~Mistlberger,
\newblock \emph{{Charged current Drell-Yan production at N$^{3}$LO}},
\newblock JHEP \textbf{11}, 143 (2020),
\newblock \doi{10.1007/JHEP11(2020)143},
\newblock \eprint{2007.13313}.

\bibitem{Camarda:2021ict}
S.~Camarda, L.~Cieri and G.~Ferrera,
\newblock \emph{{Drell-Yan lepton-pair production: $q_T$ resummation at N$^3$LL
  accuracy and fiducial cross sections at N$^3$LO}}  (2021),
\newblock \eprint{2103.04974}.

\bibitem{Dittmaier:2014qza}
S.~Dittmaier, A.~Huss and C.~Schwinn,
\newblock \emph{{Mixed QCD-electroweak $\mathcal{O}(\alpha_s\alpha)$
  corrections to Drell-Yan processes in the resonance region: pole
  approximation and non-factorizable corrections}},
\newblock Nucl. Phys. B \textbf{885}, 318 (2014),
\newblock \doi{10.1016/j.nuclphysb.2014.05.027},
\newblock \eprint{1403.3216}.

\bibitem{Catani:2007vq}
S.~Catani and M.~Grazzini,
\newblock \emph{{An NNLO subtraction formalism in hadron collisions and its
  application to Higgs boson production at the LHC}},
\newblock Phys. Rev. Lett. \textbf{98}, 222002 (2007),
\newblock \doi{10.1103/PhysRevLett.98.222002},
\newblock \eprint{hep-ph/0703012}.

\bibitem{Buonocore:2019puv}
L.~Buonocore, M.~Grazzini and F.~Tramontano,
\newblock \emph{{The $q_T$ subtraction method: electroweak corrections and
  power suppressed contributions}},
\newblock Eur. Phys. J. C \textbf{80}(3), 254 (2020),
\newblock \doi{10.1140/epjc/s10052-020-7815-z},
\newblock \eprint{1911.10166}.

\bibitem{Buonocore:2021rxx}
L.~Buonocore, M.~Grazzini, S.~Kallweit, C.~Savoini and F.~Tramontano,
\newblock \emph{{Mixed QCD-EW corrections to
  $\boldsymbol{pp\!\to\!\ell\nu_\ell\!+\!X}$ at the LHC}},
\newblock Phys. Rev. D \textbf{103}, 114012 (2021),
\newblock \doi{10.1103/PhysRevD.103.114012},
\newblock \eprint{2102.12539}.

\bibitem{Catani:2019iny}
S.~Catani, S.~Devoto, M.~Grazzini, S.~Kallweit, J.~Mazzitelli and H.~Sargsyan,
\newblock \emph{{Top-quark pair hadroproduction at next-to-next-to-leading
  order in QCD}},
\newblock Phys. Rev. D \textbf{99}(5), 051501 (2019),
\newblock \doi{10.1103/PhysRevD.99.051501},
\newblock \eprint{1901.04005}.

\bibitem{Behring:2020cqi}
A.~Behring, F.~Buccioni, F.~Caola, M.~Delto, M.~Jaquier, K.~Melnikov and
  R.~R\"ontsch,
\newblock \emph{{Mixed QCD-electroweak corrections to W-boson production in
  hadron collisions}}  (2020),
\newblock \eprint{2009.10386}.

\bibitem{Bonciani:2021zzf}
R.~Bonciani, L.~Buonocore, M.~Grazzini, S.~Kallweit, N.~Rana, F.~Tramontano and
  A.~Vicini,
\newblock \emph{{Mixed strong$-$electroweak corrections to the Drell$-$Yan
  process}}  (2021),
\newblock \eprint{2106.11953}.

\bibitem{Cascioli:2011va}
F.~Cascioli, P.~Maierh{\"o}fer and S.~Pozzorini,
\newblock \emph{{Scattering Amplitudes with Open Loops}},
\newblock Phys. Rev. Lett. \textbf{108}, 111601 (2012),
\newblock \doi{10.1103/PhysRevLett.108.111601},
\newblock \eprint{1111.5206}.

\bibitem{Buccioni:2017yxi}
F.~Buccioni, S.~Pozzorini and M.~Zoller,
\newblock \emph{{On-the-fly reduction of open loops}},
\newblock Eur. Phys. J. C \textbf{78}(1), 70 (2018),
\newblock \doi{10.1140/epjc/s10052-018-5562-1},
\newblock \eprint{1710.11452}.

\bibitem{Buccioni:2019sur}
F.~Buccioni, J.-N. Lang, J.~M. Lindert, P.~Maierh{\"o}fer, S.~Pozzorini,
  H.~Zhang and M.~F. Zoller,
\newblock \emph{{OpenLoops 2}},
\newblock Eur. Phys. J. C \textbf{79}(10), 866 (2019),
\newblock \doi{10.1140/epjc/s10052-019-7306-2},
\newblock \eprint{1907.13071}.

\bibitem{Actis:2016mpe}
S.~Actis, A.~Denner, L.~Hofer, J.-N. Lang, A.~Scharf and S.~Uccirati,
\newblock \emph{{RECOLA: REcursive Computation of One-Loop Amplitudes}},
\newblock Comput. Phys. Commun. \textbf{214}, 140 (2017),
\newblock \doi{10.1016/j.cpc.2017.01.004},
\newblock \eprint{1605.01090}.

\bibitem{Denner:2017wsf}
A.~Denner, J.-N. Lang and S.~Uccirati,
\newblock \emph{{Recola2: REcursive Computation of One-Loop Amplitudes 2}},
\newblock Comput. Phys. Commun. \textbf{224}, 346 (2018),
\newblock \doi{10.1016/j.cpc.2017.11.013},
\newblock \eprint{1711.07388}.

\bibitem{Grazzini:2017mhc}
M.~Grazzini, S.~Kallweit and M.~Wiesemann,
\newblock \emph{{Fully differential NNLO computations with MATRIX}},
\newblock Eur. Phys. J. C \textbf{78}(7), 537 (2018),
\newblock \doi{10.1140/epjc/s10052-018-5771-7},
\newblock \eprint{1711.06631}.

\bibitem{Dittmaier:2015rxo}
S.~Dittmaier, A.~Huss and C.~Schwinn,
\newblock \emph{{Dominant mixed QCD-electroweak O($\alpha$$_s$$\alpha$)
  corrections to Drell\textendash{}Yan processes in the resonance region}},
\newblock Nucl. Phys. B \textbf{904}, 216 (2016),
\newblock \doi{10.1016/j.nuclphysb.2016.01.006},
\newblock \eprint{1511.08016}.

\bibitem{Denner:2005fg}
A.~Denner, S.~Dittmaier, M.~Roth and L.~H. Wieders,
\newblock \emph{{Electroweak corrections to charged-current e+ e-
  ---\ensuremath{>} 4 fermion processes: Technical details and further
  results}},
\newblock Nucl. Phys. B \textbf{724}, 247 (2005),
\newblock \doi{10.1016/j.nuclphysb.2011.09.001},
\newblock [Erratum: Nucl.Phys.B 854, 504--507 (2012)],
\newblock \eprint{hep-ph/0505042}.

\end{thebibliography}

\nolinenumbers

\end{document}